\documentclass[aps,twocolumn,amsmath,amssymb]{revtex4-1}
\usepackage{graphicx}
\usepackage[colorlinks=true,citecolor=blue]{hyperref}

\def\vsigma{\boldsymbol{\sigma}}
\DeclareMathOperator{\Tr}{Tr}

\begin{document}

\title{Semiclassical conservation of spin and large transverse spin current in Dirac systems}

\author{Vanessa Werner}
\author{Bj\"orn Trauzettel}
\author{Oleksiy Kashuba}
\email[Email: ]{okashuba@physik.uni-wuerzburg.de}
\affiliation{Theoretische Physik IV, Institut f\"ur Theoretische Physik und Astrophysik, Universit\"at W\"urzburg, 97074 W\"urzburg, Germany}
\begin{abstract}
In Dirac materials, the low-energy excitations obey the relativistic Dirac equation.
This dependence implies that electrons are exposed to strong spin-orbit coupling.
Hence, real spin conservation is believed to be violated in Dirac materials.
We show that this point of view needs to be refined in the semiclassical picture which applies to the case of doped Dirac materials (away from the nodal point in the spectrum).
We derive a novel type of Boltzmann equation for these systems if they are brought slightly out of equilibrium.
Remarkably, spin-momentum locking is softened and a generalized spin conservation law can be formulated.
The most striking observable consequence of our theory is a large transverse spin current in a nearly ballistic transport regime.
\end{abstract}

\maketitle

\textit{Introduction.}---Recent advances in manufacturing novel materials with nodal band structure stimulated the interest of the condensed matter community to Dirac materials~\cite{Wehling2014}.
Those quantum materials host solid-state systems possessing fermion excitations with linear dispersion relation.
In particular, Dirac materials are characterized by two distinct properties of their low-energy physics: (i) strong spin-momentum locking and (ii) double-cone-shaped spectrum with both valence and conduction band touching each other at distinct points in momentum space.
Prime examples of Dirac materials are graphene in two spatial dimensions (2D)~\cite{CastroNeto2009}, the 2D surface states of 3D topological insulators (like Bi$_2$Se$_3$, tensile-strained HgTe, or $\alpha$-Sn)~\cite{Hasan2010,Qi2011}, Weyl or Dirac semimetals in 3D (like TaAs, NbP, or compressively strained HgTe)~\cite{Armitage2018}, various high-temperature superconductors with $d$-wave pairing~\cite{Sun2015}, and liquid $^{3}$He~\cite{Volovik1992}.

\begin{figure}
\centering
\includegraphics[width=.9\columnwidth,page=1]{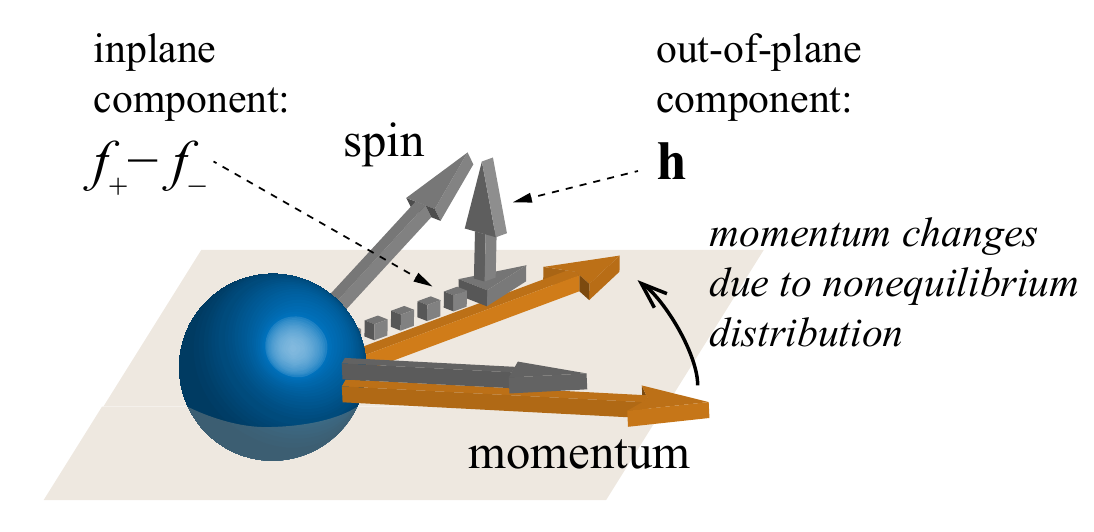}
\caption{Schematic illustration of the spin-momentum decoupling in a 2D Dirac system.
The in-plane rotation of the momentum (golden arrows) due to electric field, Lorentz force, or density gradient, causes the in-plane rotation of the spin (gray arrows) because of spin-momentum locking.
The precession of the spin in the plane, however, generates a nonzero polarization of the spin perpendicular to the plane (similar to a Larmor precession).
This results in an out-of-plane component of the spin denoted by $\mathbf{h}$.}
\label{fig:physics}
\end{figure}

Semiclassical approximations based on the Boltzmann equation have been developed to predict particular transport properties of 2D Dirac materials~\cite{Shon1998,Ando2006,Nomura2006,Adam2007,Hwang2007,Nomura2007,Auslender2007,Katsnelson2008,Fritz2008,Muller2008b,Kashuba2018}.
In the semiclassical treatment, the Dirac nature of the excitations manifests itself, for instance, in two distribution functions for each of the two Dirac cones~\cite{Shon1998,Auslender2007}, the renormalization of the mean free path due to disorder~\cite{Ando2006,Adam2007,Hwang2007,Kechedzhi2008,Katsnelson2008}, and a distinct relaxation rate in the collision integral describing  electron-electron scattering~\cite{Ando2006,Fritz2008,Muller2008b}.
The bare spin degree of freedom is generally assumed to be irrelevant due to strong spin-momentum locking.
It is described by the helicity which is fixed within a single Dirac cone.
Concretely, it is $+1$ for the top cone, the conduction band, where spin and momentum are collinear (see Fig.~\ref{fig:physics}), and it is $-1$ for the bottom cone, the valence band, where spin and momentum are anticollinear.
Such point of view, however, does not allow to take into account the deflection of the spin from the direction of motion, which requires a more elaborate treatment.
To resolve this issue for similar effects in strong spin-orbit-coupled materials, the wave package approach was developed~\cite{Niu1999,Niu2004,Niu2006}, which allows us to take into account the torque dipole contribution.
In this Letter, we propose a different formalism that properly connects the matrix structure of the Dirac equation with semiclassical transport theory without using an adiabatic approximation.

On the basis of quantum kinetic equations, it is possible to describe the kinetics of the fermion excitations in the presence of external potentials, disorder, weak interaction, etc.~\cite{RammerSmith1986,Rammer2007}.
This approach allows us to formally derive a quantum Boltzmann equation that takes into account a deviation from perfect spin-momentum locking (as illustrated in Fig.~\ref{fig:physics}).
Below, we demonstrate that this extension of the theory is important for the correct semiclassical description of Dirac materials.
It turns out that it fundamentally affects the proper prediction of spin transport in these systems.
In fact, we show below that, in the semiclassical regime, where the Fermi wavelength is small compared to other length scales in the problem, a novel type of spin conservation law holds with an elaborated definition of the semiclassical spin current.
This spin current that we predict on the basis of a refined matrix kinetic equation turns out to be quite large, as compared to spin currents in conventional (non-Dirac) materials~\cite{Wang2014,Sun2016}.

\textit{Kinetic equation.}---In order to study the nonequilibrium dynamics of electrons in Dirac materials, we first derive a kinetic equation for the semiclassical propagator for the ideally free case with no interactions present.
Note that, due to the noninteracting Hamiltonian being linear in momentum, this equation does not need a gradient approximation~\footnote{%
The triagonal warping, often present in known Dirac materials, is described by higher order terms in the momentum expansion.
Thus, the semiclassical description for warped spectra does require the gradient approximation.
The latter is valid for large Fermi momenta, but since the triagonal warping is generally relevant away from the vicinity of the Dirac point, it does not hinder the applicability of the semiclassical approach.%
}.
The second quantized electron ladder operator is a spinor that depends on time and coordinate $\psi\equiv \psi_{\alpha}(t,\mathbf{r})$ and obeys the Dirac equations
\begin{equation}
i\partial_{t}\psi = -i\vsigma\cdot\nabla\psi,
\qquad
-i\partial_{t}\psi^{+} = i(\nabla\psi^{+})\cdot\vsigma,
\label{eq:schroe}
\end{equation}
where $\vsigma$ is a vector of three (in 3D) and two (in 2D) Pauli matrices.
We put the Fermi velocity $v$ and $\hbar$ to unity in order to simplify the formulas, but we will restore natural units when we present the final results.
Let us introduce the semiclassical Green function~\footnote{%
We use the term ``semiclassical'' Green function since the ladder operators are taken at the same time.
This is equivalent to the integration of the lesser Green function over frequency.
In the noninteracting case, the spectral function is almost a delta function.
This means that by integrating over frequency, we put the Green function on the mass surface excluding virtual processes.}
\begin{equation}
\!\!g_{\alpha\beta}(t,\mathbf{r},\mathbf{p}) \!=\!\!\! \int \!\!\left\langle \!\psi^{+}_{\beta}\! \left(t,\mathbf{r}\!+\!\frac{\mathbf{a}}{2}\right)\!\psi_{\alpha}\!\left(t,\mathbf{r}\!-\!\frac{\mathbf{a}}{2}\right)\!\right\rangle \! e^{-i\mathbf{p}\mathbf{a}} d^{D}\!\mathbf{a},
\end{equation}
where $D=2,3$ is the dimensionality and the integral is taken over the volume.
This Green function allows us to calculate local observables.
It can be understood as a generalized version of the distribution function.
For electron density and current, we obtain, respectively
\begin{align}
\varrho(t,\mathbf{r}) &= \sum_{\mathbf{p}} \Tr[g(t,\mathbf{r},\mathbf{p})],
\\
\mathbf{j}(t,\mathbf{r}) &= \sum_{\mathbf{p}} \Tr[\vsigma g(t,\mathbf{r},\mathbf{p})], \label{jt}
\end{align}
where $\sum_{\mathbf{p}}=\int\frac{d^{D}\mathbf{p}}{(2\pi)^{D}}$, and the elementary charge is also put to unity.
In dimensionless units, putting $\hbar \equiv e \equiv v \equiv 1$, the expression for the current $\mathbf{j}(t,\mathbf{r})$, Eq.~(\ref{jt}), coincides with the spin density $\mathbf{s}(t,\mathbf{r})$ in 3D because of spin-momentum locking in the Dirac equation.
In 2D, however, this coincidence is true only for the $x$ and $y$ components, since the current and momenta are confined to a plane.
Importantly, the spin can also have an out-of-plane component with the notation $s_{z}= \sum_{\mathbf{p}} \Tr[\sigma^{z} g]$ in the 2D case.
On the basis of Eq.~\eqref{eq:schroe}, we derive the matrix kinetic equation
\begin{equation}
\partial_{t}g = -\frac12 \nabla \cdot \{\vsigma , g \}_{+} - i\mathbf{p} \cdot [\vsigma,g]_{-},
\label{eq:gdynamics}
\end{equation}
which is exact in the clean and noninteracting case.

In equilibrium, the expressions for the semiclassical Green functions are well known and are given by
\begin{equation}
g_\text{eq} = \sum_{\lambda=\pm}\frac{1+\lambda\mathbf{n}\cdot\vsigma}{2} f_{F}(\lambda p),
\label{eq:gequilibrium}
\end{equation}
where $\mathbf{n}=\mathbf{p}/|\mathbf{p}|$, $f_{F}(\varepsilon)=1/(e^{(\varepsilon-\mu)/T}+1)$ is the Fermi distribution function, with $\mu$ being the chemical potential, and $T$ the temperature of the electron system.
In a more general case, however, we can parametrize the function $g$ by the distribution functions  $f_+$ and $f_-$ corresponding to the positive and negative helicities, respectively, and the transverse amplitude $\mathbf{h}$, a vector perpendicular to momentum, i.e., \mbox{$\mathbf{h}\cdot\mathbf{n}=0$} (see Fig.~\ref{fig:physics}),
\begin{equation}
g = \sum_{\lambda=\pm}\frac{1+\lambda\mathbf{n}\cdot\vsigma}{2} f_{\lambda} + \vsigma\cdot\mathbf{h}.
\label{eq:gansatz}
\end{equation}
In this parametrization we can always choose $\mathbf{h}\perp \mathbf{n}$, attributing the components parallel to the momentum to $f_{+}$ and $f_{-}$.
Note that here $f_{+}$ and $f_{-}$ are not exactly the distributions in the top and bottom cones.
The function $g$ is a function of coordinate $\mathbf{r}$ and momentum $\mathbf{p}$, but not energy.
Therefore, it partially mixes top and bottom cone states.
Because of the deflection of the spin from the momentum, say in the top cone, the spin may also have components perpendicular to momentum described by $\mathbf{h}$, and anticollinear to it, described by $f_{-}$.
Hence, the spin component parallel to momentum is $f_{+}-f_{-}$, as shown in Fig.~\ref{fig:physics}.
In the uniform and stationary case, the $f_{-}$ distribution function vanishes for the top cone , while $f_{+}$ becomes the conventional distribution function.
In the bottom cone, $f_{+}$ and $f_{-}$ exchange roles.

Substituting this ansatz into Eq.~\eqref{eq:gdynamics}, we obtain the following set of equations (see Appendix~\ref{SuppMatA})
\begin{align}
\dot{f}_{\pm} \pm \mathbf{n}\cdot \nabla f_{\pm} + \nabla \cdot \mathbf{h} &= 0,
\label{eq:dynamicsf}
\\
\dot{\mathbf{h}} - 2\mathbf{p}\times\mathbf{h} - \frac12\mathbf{n}\times\bigl[\mathbf{n}\times\nabla(f_{+}+f_{-})\bigr] &= 0.
\label{eq:dynamicsh}
\end{align}
The central technical point of our paper is reflected in the role of $\mathbf{h}$ in these equations.
In equilibrium, we expect that $\mathbf{h}=0$, as follows from Eq.~\eqref{eq:gequilibrium}.
Furthermore, if $\nabla f_{\pm}=0$, Eq.~\eqref{eq:dynamicsh} reduces to $\dot{\mathbf{h}} = 2\mathbf{p}\times\mathbf{h}$, which allows for a solution of $\mathbf{h}$ in the form of a vector of arbitrary length perpendicular to $\mathbf{p}$ rotating with frequency $2p$ around it, $\mathbf{h}(t)=\mathrm{Re}[e^{2ipt}\mathbf{h}(0)]$, where $p=|\mathbf{p}|$.
This solution is quickly oscillating away from the Dirac point.

Without the assumption $\nabla f_{\pm}=0$, the full set of equations~\eqref{eq:dynamicsf} and~\eqref{eq:dynamicsh} is characterized by two types of frequencies: (i) proportional to the gradients of the distribution functions $\nabla f_{\pm}$ and (ii) proportional to $2p$ (see Appendix~\ref{SuppMatB}).
We are particularly interested in a semiclassical picture, where the Fermi wavelength of the electrons is small compared to other length scales of the problem.
Then, the frequencies of type (i) are much smaller than the ones of type (ii).
Furthermore, in the semiclassical approximation, quantum coherence is lost due to relaxation.
Notably, the parameter $\mathbf{h}$ describes the dynamics of the (off-diagonal) elements of the spin density matrix.
All of the off-diagonal elements are subjected to relaxation due to interaction of the system with its environment.
However, the boundary conditions for $f_{\pm}$ play the role of a source exciting the low-frequency modes proportional to the gradients.
This is a particular property of our refined Dirac semiclassics.
Under these conditions, the high-frequency modes vanish due to relaxation if the momentum $p$ is much larger than the typical scales of the distribution function gradients.

Thus, in the semiclassical picture, we look for solutions that behave as $\mathbf{h}\to0$ if $\nabla f_{\pm}\to0$.
This requirement applied to Eq.~\eqref{eq:dynamicsh} implies that additionally $\mathbf{h}\to0$ in the limit of $p\to\infty$.
These two observations guide us to make the ansatz
\begin{equation}
\mathbf{h} = \frac{1}{4p}\nabla\check{H},
\quad \check{H}\cdot\mathbf{n}=0,\quad
(\check{H})_{ij}\equiv H_{ij}[f_{\pm}].
\label{eq:hdH}
\end{equation}
In these equations, the divergence and the scalar product should be understood as $(\nabla \check{H})_{i}=\sum_{j}\nabla_{j}H_{ji}$ and $(\check{H}\cdot\mathbf{n})_{i}=\sum_{j}H_{ij}\mathbf{n}_{j}$, where $j=x,y,z$, while $i=x,y$ in 2D, and $i=x,y,z$ in 3D.
The last part of Eq.~\eqref{eq:hdH} stresses that the matrix $\check{H}$ is a functional of the distribution functions $f_{\pm}$.

Expanding in $1/p$ and keeping all terms up to first order, we derive an explicit expression for the generalized distribution function
\begin{equation}
g \approx \sum_{\lambda=\pm}\left[\frac{1+\lambda\mathbf{n}\cdot\vsigma}{2} f_{\lambda} - \frac{1}{4p}\vsigma\cdot(\mathbf{n}\times\nabla f_{\lambda})\right].
\label{eq:gapprox}
\end{equation}
The functions $f_{\lambda}$ obey the standard Boltzmann equations (see Appendix~\ref{SuppMatA}):
\begin{equation}
\partial_{t}f_{\pm} \pm \mathbf{n}\cdot \nabla f_{\pm} = 0.
\end{equation}
Note that Eq.~\eqref{eq:gapprox} is written for the 3D case, which could, for example, apply to Weyl semimetals.
In 2D, corresponding to surface states of 3D TIs or graphene, the solution takes a slightly different form
\begin{equation}
g \approx \sum_{\lambda=\pm}\left[\frac{1+\lambda\mathbf{n}\cdot\vsigma}{2} f_{\lambda} - \frac{\sigma_{z}}{4p}(\mathbf{n}\times\nabla f_{\lambda})_{z}\right],
\label{eq:gapprox2D}
\end{equation}
which implies that the spin polarization acquires an out-of-plane component (see Fig.~\ref{fig:physics}), in the presence of a spatial variation of $f_{\pm}$.

\textit{Conservation laws.}---Before we investigate the spin dynamics, let us consider particle conservation first.
Taking the trace of Eq.~\eqref{eq:gdynamics}, we derive the charge conservation law
\begin{equation}
\partial_{t}\varrho + \nabla\cdot\mathbf{j} =0.
\label{eq:conservecharge}
\end{equation}
This law is exact; i.e., it is correct in all orders of the $p\gg \nabla$ expansion.
Taking the trace with the Pauli matrix multiplied to Eq.~\eqref{eq:gdynamics}, we derive the expression for the time derivative of the current density.
In dimensionless units, it is equivalent to the time derivative of the spin density given by
\begin{equation}
\partial_{t}\mathbf{j} \!=\!
- \nabla \varrho - 2\sum_{\mathbf{p}}\Tr[(\vsigma\times\mathbf{p})g]
\!=\!
- \nabla \varrho + 4 \sum_{\mathbf{p}}\mathbf{p}\times \mathbf{h}.
\label{eq:directcs}
\end{equation}
We may notice that the term breaking the conservation of the spin is the antisymmetric part of the stress-energy tensor, which for the Hamiltonian~\eqref{eq:schroe} is defined as $\langle\psi^{+}p^{i}\sigma^{j}\psi\rangle$.
This follows from both momentum and total angular momentum conservation laws (see Appendix~\ref{SuppMatD} and the Ref.~\cite{Belinfante1939}).

At first glance, the conservation of spin seems to be broken in Eq.~\eqref{eq:directcs}, which is expected for systems with strong spin-orbit coupling.
However, if we assume perfect spin-momentum locking, as most of the works on semiclassics do, the vector product $\vsigma\times\mathbf{p}$ vanishes.
Interestingly, on the basis of our refined Dirac semiclassical approximation, it is possible to derive a novel type of spin conservation with an emerging spin current.
Using the ansatz~\eqref{eq:hdH}, we define an effective spin current $\check{\Pi}$ in the following way:
\begin{equation}
\partial_{t}\mathbf{j} + \nabla \check{\Pi}=0,
\qquad
\check{\Pi} = \varrho \check{1} + \sum_{\mathbf{p}} \check{H}\times\mathbf{n}.
\label{eq:conservespin}
\end{equation}
Each element $\Pi_{ij}$ of the spin current specifies the flow of the spin component $j$ in the direction $i$.
Here, $\check{1}$ stands for $\delta_{ij}$ and the vector product in the second term is understood as $\sum_{kl} \varepsilon_{lkj} H_{il}n_{k}$.
Remember that, in Dirac materials, the current density~$\mathbf{j}$ is equivalent to the spin density~$\mathbf{s}$.

To obtain an explicit expression for the spin current $\check{\Pi}$, we calculate it using the approximation~\eqref{eq:gapprox} for the 3D case.
Then, already to zeroth order in $1/p$, we obtain for charge, charge current or spin, and spin current densities (see Appendix~\ref{SuppMatC})
\begin{equation}
\varrho= \sum_{\pm,\mathbf{p}} f_{\pm},
\quad
\mathbf{j}\approx \sum_{\pm,\mathbf{p}}(\pm\mathbf{n}) f_{\pm},
\quad
\Pi_{ij}\approx\sum_{\pm,\mathbf{p}}  n_{i}n_{j}f_{\pm} .
\label{eq:jPiapprox}
\end{equation}
In 2D, the gradient and the momentum have only $x$ and $y$ components, but we additionally have the out-of-plane spin polarization in first order of $1/p$  described by Eq.~\eqref{eq:gapprox2D}.
The spin flow connected with the conservation of the out-of-plane spin is small by $1/p$ (see Appendix~\ref{SuppMatC}).
The in-plane spin conservation law for 2D obeys Eq.~\eqref{eq:conservespin} and is described by Eq.~\eqref{eq:jPiapprox} where the indices run over the in-plane coordinates only, i.e.\ $i,j=x,y$.

We would like to stress an important fact concerning the contribution to the spin current originating from the transverse component $\mathbf{h}$.
Despite the fact that this component is of first order in $1/p$, it contributes to zeroth order to the spin current.
This can easily be seen from Eq.~\eqref{eq:directcs} where $\mathbf{h}$ is multiplied by $\mathbf{p}$ in the second term on the right-hand side.
Hence, the influence of $\mathbf{h}$ on the effective spin current is not small.


%
\begin{figure}
\centering
\includegraphics[width=.75\columnwidth,page=1]{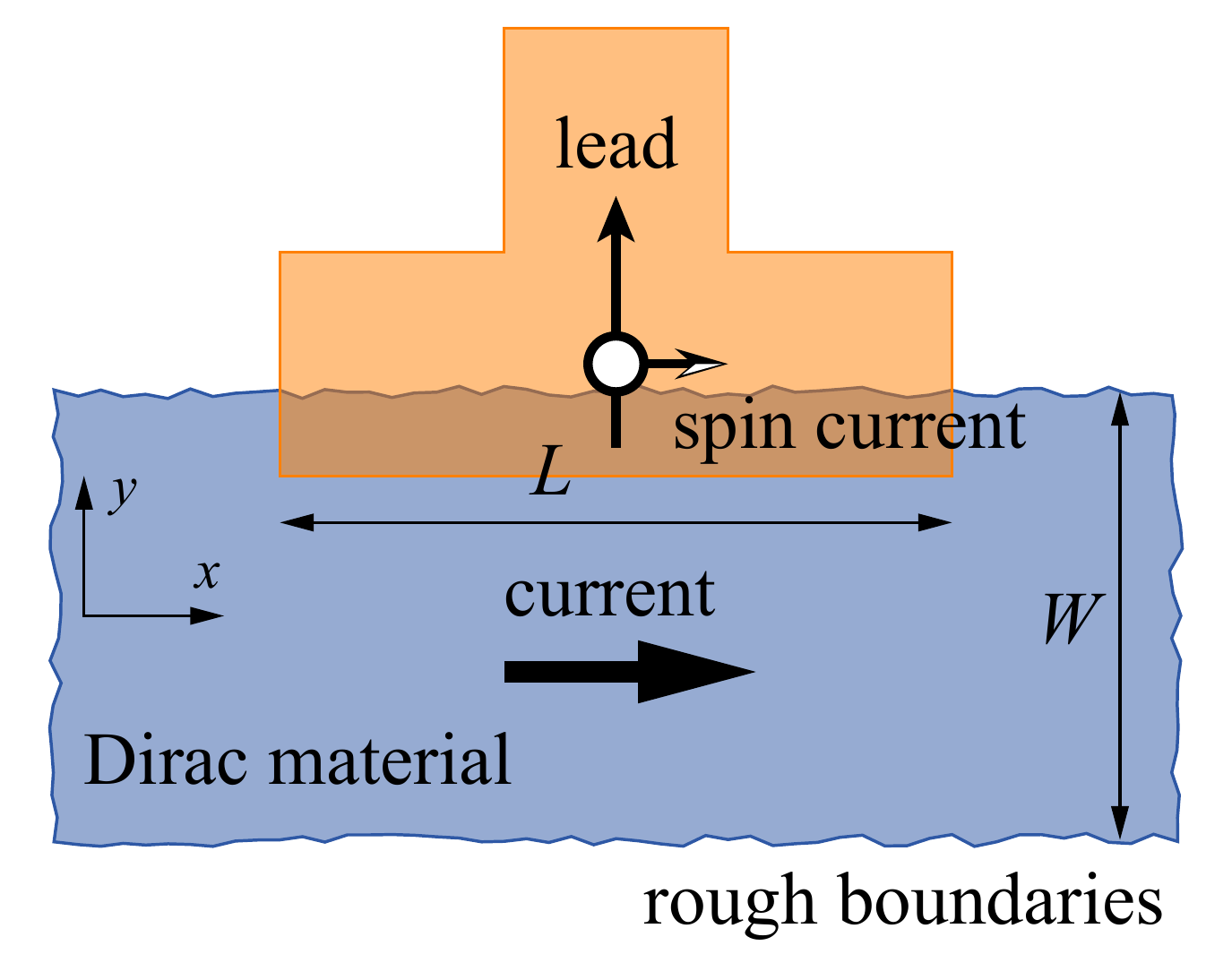}
\caption{Experimental setup for collecting the spin current from a 2D Dirac channel at the surface of a 3D topological insulator.}
\label{fig:setup}
\end{figure}
\textit{Experimental consequences.}---The theory stated above allows its direct application in a topological insulator setup.
Unlike graphene, the spin operator in the corresponding Dirac equation describes the real electron spin.
Therefore, it remains well defined outside the Dirac material, in particular, in a metal with weak spin-orbit interaction.
This means that the spin current in the Dirac sample can be injected into an attached lead as schematically shown in Fig.~\ref{fig:setup}.
Let us try to estimate the magnitude of this effect.

In our calculation, we assume that the distribution function is close to equilibrium, meaning \mbox{$f_{\pm}= f_F(\pm vp - \chi)$}, where $v$ is the Fermi velocity, and $\chi$ is a small non-equilibrium correction giving rise to charge transport~\footnote{%
The theory is also applicable for thermotransport, when the temperature gradients instead of the bias are present.
The expression for the transverse spin current $\Delta\Pi$ in Eq.~\eqref{eq:DeltaPiballistic} remains, while in the expression for the current density $j_{x}$, the effective electric field $E$ should be substituted by the temperature gradient times the Seebeck coefficient.%
}.
Since we use the free kinetic equation without external potential, i.e., Eq.~\eqref{eq:gdynamics}, we describe the applied bias by a gradient of the chemical potential $\partial_{x}\mu = -eE$ instead of an external electric field~\footnote{%
The implementation of the external potential in the Boltzmann equation is based on the fact that the solution for the spectral function in gradient approximation can be easily obtained from the equilibrium expression by adding the potential to the dispersion relation~\cite{Rammer2007}.
This ansatz fails for systems with strong spin-orbit coupling.
Therefore the gradients of chemical and external potentials in such systems are not equivalent anymore.%
}.
The trivial contribution to the spin current density (proportional to the unit matrix in spin space) comes from averaging the Fermi distribution over momentum direction $\mathbf{n}$.
It is proportional to the charge density $\varrho$~\footnote{%
Note the prefactor $1/2$ in the first term, contrary to the case when the spin deflection is neglected, i.e.\ $H$ is set to zero in Eq.~\eqref{eq:conservespin}.}:
\begin{equation}
\Pi_{ij} = \frac{v}{2}\varrho \delta_{ij} + \Delta\Pi_{ij}.
\label{eq:DeltaPidef}
\end{equation}
Additionally, there is a correction $\Delta\check{\Pi}$, which gives rise to new physics.
We first calculate it in the leading order of $1/p_{F}=v/|\mu|$ for the ballistic case.
When the chemical potential gradient $E$ is directed along the channel, the first nonvanishing contribution to the spin current $\Delta\check{\Pi}$ arises already in the first order of the expansion $f \approx f_F - \partial_\varepsilon f_F \chi$ and reads (see Appendix~\ref{SuppMatE})
\begin{equation}
\Delta\check{\Pi}_\text{bal} = \frac{\pi}{2}\frac{j_{x}}{e} \!\left(\! \frac yW \!-\! \frac12 \!\right)\!
\begin{pmatrix}
0 & 1 \\ 1 & 0
\end{pmatrix},
\quad
j_{x}\approx\frac{e^{2}Ep_{F}W}{2\pi^{2}\hbar^{2}}.
\label{eq:DeltaPiballistic}
\end{equation}
This result implies the transverse flow of the $x$ component of the spin.
In our coordinate system, the $x$ direction points along the channel and the $y$ direction transverse to it.
Attaching a lead, as is schematically shown in Fig.~\ref{fig:setup}, we are able to generate a spin current (into the lead) with magnitude $I_\text{spin}=\frac{\pi}{4e} \frac{L}{W} I$, where $L$ is the lead-sample interface length, and $I$ is the electric current through the channel.


The spin current is conserved locally but disappears at the wall.
The wall is assumed to scatter electrons randomly and, hence, to wash out the spin polarization.
If instead of the wall we allow the spin flow to proceed further, into an attached lead, as shown in Fig.~\ref{fig:setup}, the proposed setup can function as a generator for spin current.
This mechanism also works in the reversed way: Pumping a spin current through a lead, we can generate a current in the sample in a direction set by the spin polarization of the lead.
Remarkably, if the size of the side reservoir is not much shorter than the width of the channel, then the generated spin current is comparable to the applied electric current (up to the proper change of units).
This charge to spin conversion ratio close to unity makes our effect highly efficient.
Spin currents are generally measured either in standard Amperes, or spins per seconds.
Thus a charge current $I$ of $1\,\text{mA}$ will give approximately the same spin current in mA, which corresponds to $10^{16}$ spins per second.
The largest spin currents in conventional devices of similar size are reported to be $0.1$--$0.5$ mA~\cite{Wang2014,Sun2016}.

We have so far derived the spin current conservation in the absence of electron scattering, which can violate the spin conservation law.
This means that we need a ballistic system in order to realize the conservation law~\eqref{eq:conservespin} of the spin current~\eqref{eq:jPiapprox}.
However, the corresponding spin current can also be calculated for the cases of transport in diffusive and hydrodynamic regimes.
The adapted calculations (see Appendix~\ref{SuppMatE}) show that, in both transport regimes, there is no transverse component to the spin current, and the longitudinal component is small due to an extra factor $1/p_{F}^{2}$ at fixed current.

\textit{Conclusions.--} We have demonstrated that the kinetics of Dirac systems cannot be described solely by the dynamics of the distribution function obeying the simple Boltzmann equation.
The semiclassical approximation in Dirac systems can restore the spin conservation broken by strong spin-orbit coupling at the quantum level.
The dynamics of the spin, describing the deviation from rigid spin-momentum locking, is small by $1/p_{F}$ but plays an important role in the non-equilibrium description of the Dirac system.
The spatial variation of distribution functions in 2D Dirac systems deflect the electron spin out of the plane.
We have derived a modified spin conservation law and proposed a new definition of the semiclassical spin current.
Finally, we have estimated that our refined definition of semiclassics implies the prediction of a large spin current.

Financial support by the DFG (SPP1666 and SFB1170 "ToCoTronics") and the ENB Graduate School on Topological Insulators is gratefully acknowledged.
We thank L.W.~Molenkamp and C.~Tutschku for discussions.

\appendix
\onecolumngrid

\renewcommand\thesection{\Alph{section}}

\section{Derivation and the full solution of the kinetic equation}
\label{SuppMatA}

Let us find the solution of the Eq.~\eqref{eq:gdynamics} in the form of the ansatz in Eq.~\eqref{eq:gansatz}:
\begin{equation}
\partial_{t}g = -\frac12 \nabla \cdot \{\vsigma , g \}_{+} - i\mathbf{p} \cdot [\vsigma,g]_{-},
\qquad
g = \frac{1+\mathbf{n}\cdot\vsigma}{2}f_{+} + \frac{1-\mathbf{n}\cdot\vsigma}{2} f_{-} + \vsigma\cdot\mathbf{h}
\end{equation}
where $\mathbf{h}\cdot\mathbf{n}=0$.
Substituting the ansatz into the equation, we get
\begin{multline}
\sum_{\pm}\frac12(1\pm\mathbf{n}\cdot\vsigma) \dot{f}_{\pm} + \vsigma\cdot\dot{\mathbf{h}} =
 -\sum_{\pm}\frac12  (\pm\mathbf{n}+\vsigma)\cdot \nabla f_{\pm} - \nabla \cdot\mathbf{h} + 2(\vsigma\times\mathbf{p}) \cdot \mathbf{h} =\\=
 -\sum_{\pm}\frac12  (1\pm\vsigma\cdot\mathbf{n})\left(\pm\mathbf{n}\cdot \nabla f_{\pm} + \nabla \cdot\mathbf{h} \right)  + \sum_{\pm} \left(p \mathbf{h} + \frac12 \mathbf{n}\times\nabla f_{\pm}\right)\cdot(\vsigma\times\mathbf{n}).
\label{eq:matrixeq}
\end{multline}
This matrix equation can be split into the components multiplied to $\vsigma\times\mathbf{n}$ and the two projection operators $1\pm\mathbf{n}\cdot\vsigma$:
\begin{equation}
\dot{f}_{\pm} \pm \mathbf{n}\cdot \nabla f_{\pm} + \nabla \cdot \mathbf{h}=0,
\qquad
\dot{\mathbf{h}} - 2 \mathbf{p}\times\mathbf{h} - \frac12\mathbf{n}\times\left(\mathbf{n}\times\nabla\right)  \sum_{\pm}f_{\pm}=0.
\end{equation}
The first equation describes the dynamics of the distribution functions $f_{\pm}$, while the second one allows to express the transverse component $\mathbf{h}$ through the distribution functions $f_{\pm}$.
Assuming a large momentum in comparison to the gradients, $\mathbf{p}\gg\nabla$, we expand the solution of the second equation as $\mathbf{h}=\mathbf{h}^{(0)}+\mathbf{h}^{(1)}+\mathbf{h}^{(2)}+\ldots$ leaving in the equations the zero and first (proportional to $1/p$) order terms only:
\begin{equation}
\dot{f}_{\pm} \pm \mathbf{n}\cdot \nabla f_{\pm} + \nabla \cdot \mathbf{h}^{(1)}=0,
\qquad
\dot{\mathbf{h}}^{(1)} - 2 \mathbf{p}\times(\mathbf{h}^{(1)}+\mathbf{h}^{(2)}) - \frac12\mathbf{n}\times\left(\mathbf{n}\times\nabla\right)  \sum_{\pm}f_{\pm}=0.
\end{equation}
The zeroth order solution is by construction trivial, i.e.~$\mathbf{h}^{(0)}=0$, while the first order solution is
\begin{equation}
\mathbf{h}^{(1)} = -\frac{1}{4p} \mathbf{n}\times\nabla (f_{+}+f_{-}).
\end{equation}
Note that $\nabla \cdot \mathbf{h}^{(1)} =0$.
This fully removes the transverse component from the equations for the distribution function's dynamics, i.e.\
\begin{equation*}
\dot{f}_{\pm} \pm \mathbf{n}\cdot \nabla f_{\pm} = O[1/p^{2}].
\end{equation*}
The second order correction is
\begin{equation}
\mathbf{h}^{(2)} = -\frac{1}{2p} \mathbf{n}\times \partial_{t} \mathbf{h}^{(1)} = \frac{1}{8p^{2}} \mathbf{n}\times(\mathbf{n}\times\nabla \partial_{t} (f_{+}+f_{-}))
=
\frac{1}{8p^{2}} \mathbf{n}\times\Bigl(\mathbf{n}\times\nabla \bigl(-\mathbf{n}\cdot\nabla(f_{+}-f_{-})\bigr)\Bigr).
\end{equation}
In principle, we can derive a full solution for $\mathbf{h}$ on the basis of the recurrent relation
\begin{equation}
\mathbf{h}^{(n+1)} = - \frac{1}{2p} \mathbf{n}\times\dot{\mathbf{h}}^{(n)}.
\end{equation}
Using the expression for $\mathbf{h}^{(1)}$ and summing up all orders, we get
\begin{equation}
\mathbf{h} = -\frac{1}{4p}\left( 1+\frac{\partial_{t} }{2p}\right)^{-1}\left(\mathbf{n}\times\nabla - \frac{\partial_{t}}{2p}\mathbf{n}\times(\mathbf{n}\times\nabla)  \right)(f_{+}+f_{-}),
\end{equation}
where the inverse of the expression with a derivative should be treated by using the Taylor expansion over the derivative: $(1-\partial)^{-1}=1+\sum_{n=1}^{\infty}\partial^{n}$.
Similarly, we can write down the formal but exact expression for the matrix $\check{H}$ used in the ansatz in Eq.~\eqref{eq:hdH}:
\begin{equation}
H_{ij} = \sum_{\pm,\mathbf{p}}\left( 1+\frac{\partial_{t} }{2p}\right)^{-1}\left(-\varepsilon_{ijk}n_{k}  + (n_{i}n_{j}-\delta_{ij})\frac{\partial_{t}}{2p}  \right)f_{\pm}.
\end{equation}

\section{Stability of solutions}
\label{SuppMatB}

Since the equations are linear, let us look for the solution in the form of
\begin{equation}
\mathbf{h}(t,\mathbf{r})=\mathbf{h}e^{i\mathbf{k}\cdot\mathbf{r}-i\omega t},
\qquad
f_{\pm}(t,\mathbf{r})=f_{\pm} e^{i\mathbf{k}\cdot\mathbf{r}-i\omega t}.
\end{equation}
Splitting $\mathbf{k} = k_{||}\mathbf{n} + \mathbf{k}_{\perp}$, where $\mathbf{n} \cdot \mathbf{k}_{\perp}=0$, and $\mathbf{h} = \frac{1}{k_{\perp}}(\mathbf{k}_{\perp} h_{||}+\mathbf{n}\times\mathbf{k}_{\perp} h_{\perp})$ we get the matrix equation for the vector $X=(f_{+},f_{-},h_{||},h_{\perp})$
\begin{equation}
\Omega X =0,
\qquad\text{where}\qquad
\Omega=
\begin{pmatrix}
-\omega +  k_{||} & 0 & k_{\perp} & 0 \\
0 & -\omega -  k_{||} & k_{\perp} & 0 \\
\frac 12 k_{\perp} & \frac 12 k_{\perp} & -\omega & -2ip \\
0 & 0 & i 2p& -\omega
\end{pmatrix}.
\end{equation}
To find the resonance frequencies we calculate the determinant of the matrix $\Omega$ and obtain
\begin{equation}
0=\det\Omega = (\omega^{2}-4p^{2})(\omega^{2}-k_{||}^{2})- k_{\perp}^{2}\omega^{2}.
\end{equation}
The discriminant of this quadratic (with respect to $\omega^{2}$) equation is
\begin{equation}
\Delta = \left(4p^{2}-k_{||}^{2}\right)^{2} + 2 k_{\perp}^{2} \left(4p^{2}+k_{||}^{2}\right) + k_{\perp}^{4}
\end{equation}
and always positive.
This implies that all the roots are real and equal to
\begin{equation}
\omega =\pm \sqrt{2p^{2}+\frac12 k^{2} \pm \frac12\sqrt\Delta }.
\end{equation}
Expanding for large $p$, we get four solutions
\begin{equation}
\omega_{1/2}  \approx \pm k_{||} ,
\qquad
\omega_{3/4} \approx \pm \sqrt{4p^{2} + k_{\perp}^{2}}.
\end{equation}

\section{Antisymmetric part of the stress-energy tensor}
\label{SuppMatD}

The second term on the right-hand side of Eq.~\eqref{eq:directcs} (first equality), is the expectation value of the commutator of spin density operator with the Hamiltonian.
One can notice that for our particular case, i.e.\ the Weyl Hamiltonian, this term is identical with the antisymmetric part of the canonical stress-energy tensor $T_{ij}=\langle\psi^{+}p^{i}\sigma^{j}\psi\rangle$.
In classical field theory, according to Noether's theorem, the antisymmetric part of the canonical stress-energy tensor can always be expressed in form of the derivative of the Belinfante tensor~\cite{Belinfante1939}.
We now demonstrate how this relates to the uniformity and isotropy of the considered system, which manifests itself in the conservation of momentum and angular momentum.
The momentum density $\mathbf{P}=\sum_{\mathbf{p}}\Tr[\mathbf{p}g]$, as it follows from the Eq.~\eqref{eq:gdynamics}, is conserved:
\begin{equation}
\partial_{t}\mathbf{P}+\nabla\check{T}=0.
\end{equation}
The conservation of the total angular momentum $\mathbf{J}$, which is the sum of orbital angular momentum and spin (with the corresponding operator $\mathbf{r}\times\mathbf{p}+\frac12\vsigma$), implies the similar equation
\begin{equation}
\partial_{t}\mathbf{J}+\nabla\check{M}=0 .
\end{equation}
Furthermore, $\mathbf{J}=\mathbf{r}\times\mathbf{P}+\frac12\mathbf{j}$, which results in:
\begin{equation}
-\partial_{t}\mathbf{J}_{i}- \partial_{i}\varrho=\varepsilon_{ikl}\mathbf{r}_{k}\partial_{j}T_{jk}+\varepsilon_{ijk}T_{jk}  = \partial_{j}(\varepsilon_{ikl}\mathbf{r}_{k}T_{jl})
\end{equation}
yielding an expression for the canonical angular momentum tensor \mbox{$M_{ij}=\varepsilon_{ikl}\mathbf{r}_{k}T_{jl}+\varrho\delta_{ij}$}.
Here we used a relation $\partial_{j}\mathbf{r}_{k}=\delta_{jk}$.


\section{Spin conservation law}
\label{SuppMatC}

In order to analyze the spin conservation equation in form of Eq.~\eqref{eq:conservespin}, let us take the trace of Eq.~\eqref{eq:matrixeq} together with the Pauli matrices $\boldsymbol{\sigma}$ (but we do not integrate over the momenta):
\begin{equation}
\partial_{t}\sum_{\pm} (\pm\mathbf{n} f_{\pm} + \mathbf{h})=
\sum_{\pm}\Bigl(-\mathbf{n}(\mathbf{n}\cdot \nabla f_{\pm}) + 2p \mathbf{n}\times\mathbf{h} + \mathbf{n}\times(\mathbf{n}\times\nabla f_{\pm})\Bigr).
\end{equation}
As we see, the left part corresponds to the time derivative of the spin polarization, compare with Eq.~\eqref{eq:jPiapprox} where the expression for $\mathbf{h}^{(1)}$ is used.
In the right part, the second term contains a factor $p$.
Hence, we have to consider the second order expression for $\mathbf{h}$ to be consistent up to order $1/p$.
In the zeroth order contribution to the equation, the expression for $\mathbf{h}^{(1)}$ cancels out the third term resulting in the main contributions to current and spin current, respectively,
\begin{equation*}
\mathbf{j}^{(0)} = \sum_{\pm,\mathbf{p}}(\pm\mathbf{n}) f_{\pm},
\qquad
\nabla\check{\Pi}^{(0)} = \sum_{\pm,\mathbf{p}}\mathbf{n}(\mathbf{n}\cdot \nabla f_{\pm}).
\end{equation*}
The first order correction is $\nabla\check{\Pi}^{(1)} = -2\sum_{\pm,\mathbf{p}}\mathbf{p}\times\mathbf{h}^{(2)}$.
Using the expression for $\mathbf{h}^{(2)}$, we get
\begin{equation}
-2p\sum_{\pm}\mathbf{n}\times\mathbf{h}^{(2)}
=
-\frac{1}{2p} \mathbf{n}\times \bigl(\mathbf{n}(\mathbf{n}\cdot\nabla)-\nabla\bigr) (-\mathbf{n}\cdot\nabla)(f_{+}-f_{-})
=
-\frac{1}{2p} (\mathbf{n}\times\nabla) (\mathbf{n}\cdot\nabla)(f_{+}-f_{-}).
\end{equation}
Thus, the current and spin current require the corrections
\begin{equation*}
\mathbf{j}^{(1)}= -\sum_{\pm,\mathbf{p}}\frac{1}{2p} \mathbf{n}\times\nabla f_{\pm},
\qquad
\nabla\check{\Pi}^{(1)} = \sum_{\pm,\mathbf{p}}\frac{\mp1}{2p}(\nabla\cdot\mathbf{n})(\mathbf{n}\times\nabla) f_{\pm}.
\end{equation*}
Using the index representation, this can be rewritten as
\begin{equation}
\nabla_{j}\Pi_{ji}^{(1)}
=
\sum_{\pm,\mathbf{p}}
\frac{\mp 1}{2p} \nabla_{j}\nabla_{j'}(\varepsilon_{ijk}\delta_{k'j'})n_{k}n_{k'}f_{\pm}.
\end{equation}
Then, up to first order in $1/p$, we obtain for current and effective spin current densities
\begin{equation}
\mathbf{j}\approx \sum_{\pm,\mathbf{p}}\left(\pm\mathbf{n} f_{\pm} - \frac{1}{2p} \mathbf{n}\times\nabla f_{\pm} \right),
\qquad
\Pi_{ij}\approx\sum_{\pm,\mathbf{p}} \left( n_{i}n_{j}f_{\pm} \pm \frac{1}{2p}(\nabla\times\mathbf{n})_{i}n_{j}f_{\pm}\right).
\end{equation}
In 2D, the gradient and momentum have only $x$ and $y$ components, but we additionally have the out-of-plane spin polarization $s_{z}$.
The expression for the charge density remains the same, while the current and out-of-plane spin polarization are
\begin{equation}
\mathbf{j}\approx \sum_{\pm,\mathbf{p}}(\pm\mathbf{n}) f_{\pm},
\qquad
s_{z}\approx -\sum_{\pm,\mathbf{p}}\frac{1}{2p} (\mathbf{n}\times\nabla)_{z} f_{\pm}.
\end{equation}
The in-plane spin conservation law obeys Eq.~\eqref{eq:conservespin}, with the in-plane spin current
\begin{equation}
\Pi_{ij}\approx\sum_{\pm,\mathbf{p}} n_{i}n_{j}f_{\pm},
\qquad
i,j=x,y.
\end{equation}
For the out-of-plane spin, the conservation law reads
\begin{equation}
\partial_{t}s_{z} + \nabla \boldsymbol{\Pi}=0,
\qquad
\boldsymbol{\Pi}  \approx\sum_{\pm,\mathbf{p}} \frac{\pm \mathbf{n}}{2p}(\mathbf{n}\times\nabla)_{z}f_{\pm}.
\end{equation}
Evidently, $\boldsymbol{\Pi}$ is a vector describing the flow of the $z$-component (out-of-plane) of spin.

\section{Calculation of the spin currents}
\label{SuppMatE}

We now describe the calculation of the spin currents in three distinct transport regimes: (a) diffusive, (b) hydrodynamic, and (c) ballistic.
The last case (c) turns out to be the most interesting one.

\textit{Diffusive case.}---%
Solving the Boltzmann equation for the diffusive case, with a bias applied in $x$-direction (along the channel), the function $\chi$ is~\cite{Kashuba2018}
\begin{equation}
\chi =  e E l \cos \varphi,
\label{eq:chidiffusive}
\end{equation}
where $l$ is a mean free path generated by the disorder scattering.
Performing the expansion up to second order $f \approx f_F - \partial_\varepsilon f_F \chi + \frac12 \partial_\varepsilon^{2} f_F \chi^{2}$, for the first non-vanishing contribution to the $\Delta\check{\Pi}$, we get
\begin{equation}
\Delta\check{\Pi}_\text{diff}
=
\frac{\pi}{2v}\left(\frac{\hbar j_{x}}{e p_{F}} \right)^{2}
\begin{pmatrix}
1 & 0 \\ 0 & -1
\end{pmatrix},
\qquad
j_{x}=\frac{e^{2}E l p_{F}}{4\pi\hbar^{2}}.
\label{eq:Pidiffusive}
\end{equation}
Here $j_{x}$ is the current density and $p_{F}=|\mu|/v$ is the Fermi momentum.
Evidently, in such a case, there is no transverse spin flow across the channel.

\textit{Hydrodynamic case.}---%
Solving the kinetic equation in the hydrodynamic regime~\cite{Kashuba2018}, we can demonstrate that the distribution function for the Poiseuille flow has the same angular dependence on the momentum direction as in the diffusive case, while the drift velocity is not constant, but is coordinate dependent, following the parabolic laminar profile.
The non-equilibrium correction $\chi$ in the Poiseuille regime can be presented in the form of Eq.~\eqref{eq:chidiffusive}, where the mean free path depends on the coordinate as $2y (W-y)/l_{ee}$ with $l_{ee}$ being the electron-electron interaction length and $W$ being the width of the channel.
Then, the spin current dependence on the current density is the same as in Eq.~\eqref{eq:Pidiffusive}, while the expression for the current density reads
\begin{equation}
\Delta\check{\Pi}_\text{hydro} = \Delta\check{\Pi}_\text{diff},
\qquad
j_{x}=\frac{e^{2}E p_{F}}{2\pi\hbar^{2}}\frac{y (W-y)}{l_{ee}}.
\end{equation}
Thus, the spin flow across the channel is absent here, too.

\textit{Ballistic case.}---%
In the ballistic case, when the electric field $E$ is directed along the $x$-axis, i.e.\ along the channel, the function $\chi$ is given by the formula
\begin{equation}
\chi = e E \cot \varphi
\begin{cases}
y  &0 < \varphi < \pi, \\
y-W  &\pi < \varphi < 2 \pi.
\end{cases}
\end{equation}
The first non-vanishing contribution to the spin current $\Delta\check{\Pi}$ already arises in the first order of the expansion $f \approx f_F - \partial_\varepsilon f_F \chi$ and reads
\begin{equation}
\Delta\check{\Pi}_\text{bal} = \frac{\pi}{2}\frac{j_{x}}{e} \!\left(\! \frac yW \!-\! \frac12 \!\right)\!
\begin{pmatrix}
0 & 1 \\ 1 & 0
\end{pmatrix},
\qquad
j_{x}=\frac{e^{2}Ep_{F}W}{2\pi^{2}\hbar^{2}}.
\end{equation}
In the expression for the current density $j_x$, we omit a factor $\log(l/W)$, where $l$ is a next important scattering scale in the system.

This result demonstrates that the ballistic system, contrary to the diffusive or hydrodynamic case, implies a spin flow into the direction of the wall of the channel.
This behavior corresponds to a transverse flow of the spin component pointing along the channel.

Notably, the non-trivial term in the spin current $\Delta\check\Pi$ is small by an extra factor $1/p_{F}^{2}$ at same current for diffusive and hydrodynamic cases.


\twocolumngrid

\bibliographystyle{apsrev4-1}
\bibliography{diracspinc}

\end{document}